\documentclass[useAMS,usenatbib]{mn2e}
\usepackage{epsfig,amssymb}

\title[Solar occultation of 3C~279 in gamma rays]{Gamma-ray halo
around 3C~279: looking through the Sun on the 8th of October}
\author[M.~Fairbairn, T.~Rashba and S.~Troitsky]{Malcolm
Fairbairn$^{1,2}$, Timur Rashba$^{3,4,5}$ and Sergey
Troitsky$^{6}$\thanks{E-mail: st@ms2.inr.ac.ru}\\
$^1$ Physics, King's College London, Strand WC2R 2LS, London, UK\\
$^2$ CERN Theory Division, CH-1211 Geneva 23, Switzerland\\
$^3$ Max-Planck-Institut f\"ur Sonnensystemforschung,
Max-Planck-Str. 2, D-37191 Katlenburg-Lindau,
Germany\\
$^4$ Max-Planck-Institut f\"ur Physik
(Werner-Heisenberg-Institut), F\"ohringer Ring 6, D-80805 M\"unchen,
Germany\\
$^5$
Pushkov Institute of Terrestrial Magnetism, Ionosphere and Radiowave
Propagation (IZMIRAN), Troitsk, Moscow Region, 142190, Russia\\
$^6$ Institute for Nuclear Research of the Russian Academy of
Sciences,
60th October Anniversary prosp. 7a, 117312, Moscow, Russia
}
\begin{document}
\date{}
\pagerange{\pageref{firstpage}--\pageref{lastpage}} \pubyear{2008}
\maketitle
\label{firstpage}
\begin{abstract}
We discuss how the solar occultations of bright sources of
energetic gamma rays can be used to extract
non--trivial physical and astrophysical information, including the angular
size of the image when it is significantly smaller than the experiment's
angular resolution. We analyze the EGRET data and discuss prospects for
other instruments. The Fermi Gamma Ray Space Telescope will be able to
constrain the size of a possible halo around 3C~279 from observations it
makes on the 8th of October each year.
\end{abstract}

\begin{keywords}
gamma-rays: theory.
\end{keywords}

\section{Introduction}
\label{sec:intro}
The brightest source in the sky almost at any wavelength, the Sun is very
weak in high-energy ($E \gtrsim 100$~MeV) gamma rays. This property can be
used to study solar occultations of gamma-ray sources.

The width of the point-spread function (PSF) of telescopes detecting
photons at these energies is quite large, of order several degrees.
The enormous exposure of the
Fermi Gamma Ray Space Telescope (the telescope previously known as GLAST)
would partially compensate for the poor
resolution; however, it would be almost impossible to directly measure the
angular size of the image which may be smaller than the PSF width. On the
other hand, energetic gamma-ray images of distant sources may indeed have
a significant angular size due to the cascading of photons on the
background radiation and magnetic deflections of the cascade electrons and
positrons. It has long been known that one can obtain the angular size of
stars from lunar occultations, we suggest that one may determine the image
size of gamma-ray sources screened by the Sun \footnote{Interestingly, we
note that the Moon is much brighter than the Sun in this energy band
because of secondary emission from cosmic rays hitting the lunar surface,
see e.g.\ \citet{EGRET:SunMoon,Brigida:2009}.}.

The current collection of known energetic gamma-ray point sources is
scarce ($\sim 300$ sources detected by EGRET), so only a few are expected
to be on the strip on the sky such that they are screened by the Sun.
Fortunately, the brightest EGRET source identified with an
extragalactic object, 3C~279, has an ecliptic latitude of $0.2^\circ$ and
is screened by the Sun on the 8th of October each year. It is 3C~279 which is the
main subject of our discussion because, as we will see in
Sec.~\ref{sec:extended}, it represents a perfect target for this kind of study.

The simplest and most direct effect of an extended image size would be the detection of flux from the source during occultation.  Such a result could also be the signal of the transparency of the Sun to
gamma rays possible in several scenarios of new physics \citep{FRT};
however the parameter space of particular models is strongly
disfavoured by results of other experiments (e.g.\ \citet{CAST2007}).

In Sec.~\ref{sec:EGRET}, we
review our analysis of archival EGRET data of the 1991 occultation of 3C~279
during which a non-zero flux was indeed observed, although at a very low
statistical significance.

With a sensitive enough telescope, a more detailed study of the light
curve during ingress and egress would be possible. In
Sec.~\ref{sec:GLAST}, we discuss the potential of Fermi for this kind of a
study and mention sources other than 3C~279
while Sec.~\ref{sec:concl} summarizes our conclusions.

\section{Possible mechanisms for extended emission}
\label{sec:extended}

Very high energy photons interact with other photons in the source, with
photons along the path between the source and the earth and with photons
near to the sun.  These interactions result in the production of electrons
and positrons which themselves consequently interact with the gamma-ray
background leading to the developement of electromagnetic cascades which
result in a gradual decrease of the average energies of propagating
photons. Ambient magnetic fields deflect the electrons and positrons and,
as a result, the image of the source seen in gamma rays becomes extended.
This kind of extended image may be observed for distant sources which emit
very energetic photons (both the Universe and typical source environments
are transparent for gamma rays below $\sim 1$~GeV). Recently, the MAGIC
collaboration has reported~\citep{MAGIC} the detection of $E>200$~GeV emission
from 3C~279. Given the expected absorbtion on the cosmic infrared
background, this corresponds to an extremely high luminosity of the quasar
at very high energies~\citep{MAGICScience}. Moreover, according to both
the Hillas criterion~\citep{Hillas} and to bounds on the source parameters
from energy losses (see e.g.\  \citet{SourcesI}), the jets of 3C~279 might
provide the necessary conditions for the acceleration of cosmic rays (CR)
up to ultra-high energies (UHE), $E \gtrsim 10^{19}$~eV.  3C~279 therefore
seems to be an ideal candidate to search for an extended halo due to the
production of secondary photons.

Let us summarize briefly a few different scenarios
which could lead to the formation of an extended image: --

\textbf{Inverse Compton effect in the source environment ($\lesssim$
GeV).} It has been pointed out some time ago~\citep{Aharo:halo} that a
halo of \mbox{(sub-)GeV} inverse-Compton photons may form around the
source of very high energy gamma rays. The estimates of~\citet{Aharo:halo}
for the case of 3C~279 give rise to an expected angular size of the halo
of $\sim 0.2^\circ$ and a flux from the halo approximately equal to the
flux coming from the central point source.

\textbf{Synchrotron halo of UHECR sources ($\sim$GeV).}
For reasonable values of the magnetic field
($\gtrsim 10^{-9}$~G)
around a source of UHECR, synchrotron photons contribute to a
halo of angular size of a fraction of degree potentially detectable by
Fermi~\citep{Aharo:UHECR,Atoyan:2008uy}.

\textbf{Electromagnetic cascades on the intergalactic magnetic fields
($\sim 10^3$~GeV).}
Energetic photons from distant sources undergo electromagnetic cascades
when scattering off extragalactic background light, resulting in extended
images for TeV sources~\citep{NeronovSemikoz}. Even stronger cascading is
expected for secondary photons from UHECR sources (see
e.g.~\citet{Blasi}): because of their higher energies, they can scatter on
CMB photons as well as the infrared background, and the number density of
CMB photons is much higher. However, extended emission of this kind is too
weak at GeV energies to lead to a detectable
effect.

\textbf{Electromagnetic cascades on the solar radiation ($\sim 10^2$~GeV).}
The solar radiation is strongly concentrated in the optical band,
corresponding to the thermal emission of 5800~K, that is $\omega_{\rm
Sun} \sim 1$~eV. The pair production threshold is determined by
$E=m_e^2/\omega_{\rm Sun}\approx 260$~GeV. The optical depth of the solar
radiation
with respect to the pair production for 260~GeV photons tangent to the
solar surface is $\tau\sim 0.1$ and any secondary electrons and positrons
produced in this way would be isotropised by the $\sim 1$~G magnetic field
of the sun.  Because of this, the extended halo as viewed from earth would
be too weak to be observable (one may hope to detect such a halo when the
Sun passes in front of the regions with significant diffuse emission at
260~GeV if such regions exist, but without the timing signature, this
emission would be hard to detect).

Formation of the halo is a random process and therefore one expects that
the variability timescale of the halo should be determined by its physical
size (light minutes for the solar neighbourhood but millions of light
years for all other scenarios).

\begin{figure}
\begin{center}
\includegraphics[width=8.3cm]{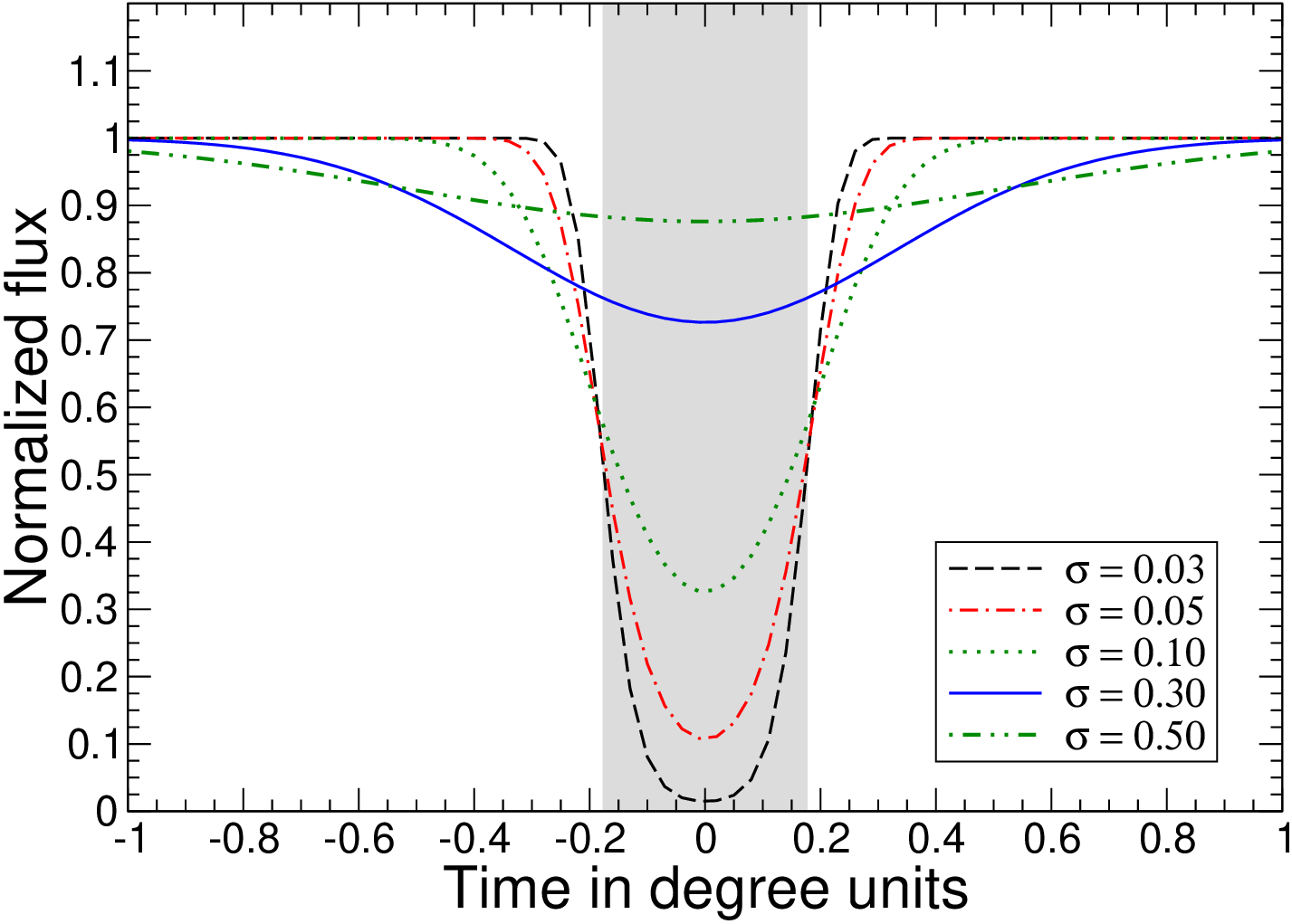}
\caption{
\label{fig:lc1}
Normalized light curves of an extended source occulted by the Sun, assuming
Gaussian model for the extended image. Different values
of the radial source variance $\sigma$ are taken in units of degree. The
time corresponds to the angle (in degree) between source center and the
point of its minimal separation from the center of the solar disk, for the
minimal separation of $0.20^\circ$.
The shadow region represents the occultation of the point-like
source.}
\end{center}
\end{figure}

We see that one might expect a halo around 3C~279 in the energy band
detectable by EGRET and Fermi (above 100~MeV). This halo
is of angular size $\sim 0.1^\circ$ while the point-spread function (PSF)
of these instruments extends for several degrees. However, the halo could
be detected when the Sun screens the bright central source. Study of the
shape and spectrum of this halo would help to distinguish between various
scenarios of its formation and therefore contribute to our understanding
of the source engine, the source environment and the extragalactic
background radiation. The extended halo would reveal itself in a smooth
falling of the flux when the Sun approaches the source and in a non-zero
flux from the source while it is screened by the Sun. In
Fig.~\ref{fig:lc1},
we present
examples of the lightcurves for various source extensions.

\section{EGRET observations}
\label{sec:EGRET}
In 1991, the solar occultation of 3C~279 occured within the field of view
of EGRET. We have analysed publicly available EGRET data to test the
conjecture that the flux from the quasar was non-zero when the source was
screened by the Sun~\citep{FRT};
here we present more details and discussions related to the study.

Given the coordinates of 3C~279 taken from
the NASA/IPAC Extragalactic Database (NED)
({\tt http://nedwww.ipac.caltech.edu})
and the
coordinates of the Sun
calculated with the program
PLANEPH~\citep{Planeph},
we are able to establish that the source was screened by the Sun for 8
hours and 34 minutes.
The minimal separation between the quasar and the center of the solar disk
was $0.20^\circ$ while the angular radius of the Sun at that period during
the earth's orbit was $0.2675^\circ$ \citep{Astrolub}. At that time
(viewing period 11.0), EGRET was pointed in the direction of
3C~273, some 5$^\circ$ away from 3C~279. The quasar was in
a moderate state and was firmly detected in gamma rays during that viewing
period~\citep{3EG}.

The actual distribution of observed photons with their coordinates,
energies and arrival times, as well as a record of the instrument's
viewing modes, are available from the EGRET ftp site
({\tt ftp://cossc.gsfc.nasa.gov/compton/data/egret}).
To calculate the exposure map for the period of occultation, we made use
of the EGRET software~\citep{FTOOLS:ftp,EGRET-software}.

To determine the source flux, one
compares the distribution of arrival directions of detected
gamma-rays with the sum of background and point-source fluxes using a
particular model for the former and the instrument point-spread function
(PSF) for the latter, both convolved with time- and direction-dependent
experimental exposure.
We denote the diffuse background flux of gamma rays from a given direction
$(\alpha ,\delta )$ as $B(\alpha ,\delta )$. This flux
was determined by \citet{EGRET:diffuse} from the analysis
of the EGRET data. From $B(\alpha ,\delta )$, we calculate $B(\psi)$, the
expected flux of background photons within the angular distance $\psi$
from the source we study (it is the integral over the corresponding circle
on the sky, of $B(\alpha ,\delta )$ weighted by exposure). The PSF of
EGRET, obtained by its careful calibration~\citep{EGRET:calibration}, is
energy dependent; one has to assume some spectral index $\alpha $ for the
source. We use the PSF for $E>100$~MeV and $\alpha=2$ (the EGRET-measured
spectral index of 3C279 is $1.96 \pm 0.04$~\citep{3EG}) from
\citet{EGRET:diffuse} and denote it as $p(\psi)$.

The number $n(\psi)$ of observed photons from the circle of angular radius
$\psi$ centered at the source (a ladder-like function of a single
variable $\psi$) is fitted by the sum of a slowly varying background and a
sharp PSF,
$$
n(\psi)=b B(\psi) +N_s p(\psi) + \sum_i s_i(\psi),
$$
where $b$ and $N_s$ are parameters of the fit (determined by
the standard least-squares method. Following the original 3EG procedure,
\citet{EGRET:like}, we keep the coefficient $b$ free motivated by possible
temporal variations of the background) and $s_i$ are contributions of
known nearby sources (for which we use the 3EG values for this viewing
period). The best-fit number of source photons $N_s$ divided by the
exposure determines the source flux. In the same way as the authors of the 3EG
catalog, we select events with energies $E>100$~MeV, TASC in coincidence
at 6~MeV and distance to the source not exceeding 15$^\circ$. These cuts
correspond also to the PSF we use.

Our approach is very similar to that used for the
construction of the 3EG catalog (see \citet{EGRET:like} for more details
of the method) apart from some details:
\begin{enumerate}
 \item
We use the updated maps of EGRET-observed diffuse
background~\citep{EGRET:diffuse} while \citet{3EG} used a theoretical
model for the distribution of the background flux.
\item
The catalog construction used the counts distribution binned in two
celestial coordinates; we use the unbinned distribution in one coordinate
-- distance from the source (both approaches were discussed by
\citet{EGRET:like}). This is more appropriate in the case of small number
of observed events.
\item
The 3EG procedure substracts all 3-sigma sources (in the catalog, they
list only 4-sigma ones). We substract only the sources listed in the
catalog, even if they are fainter than 4-sigma for this particular viewing
period (in this way, we potentially exclude some sources which
did not pass the 4-sigma cut for this viewing period, or for any other
viewing period or their sum, but were brighter than 3-sigma in this
period).
\end{enumerate}

\begin{figure}
\begin{center}
\includegraphics[width=8.3cm]{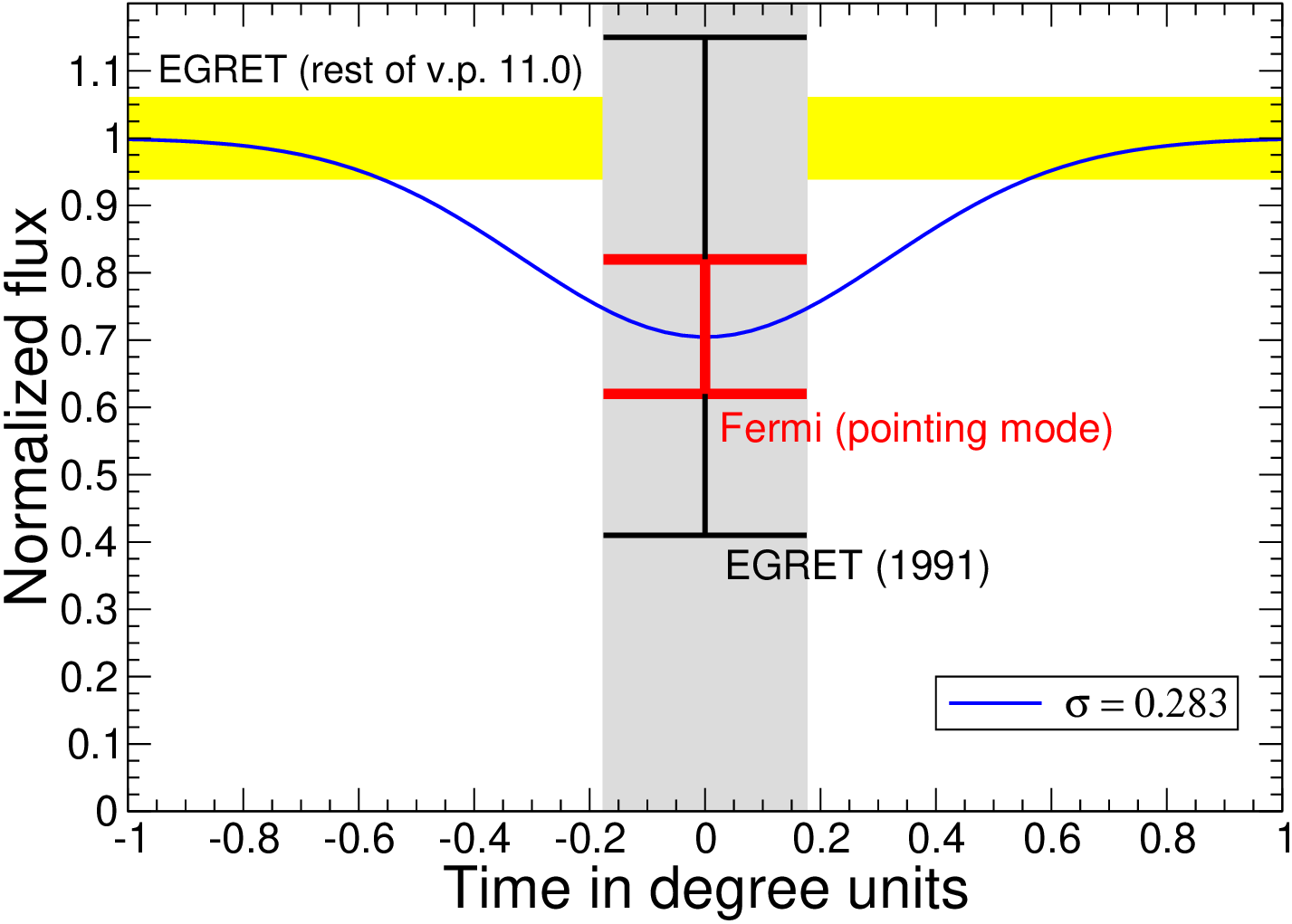}
\caption{
\label{fig:lc2}
The normalized light curve (blue line) for the Gaussian
extended source with the extension providing the best fit to the EGRET
data on 3C~279 occultation in 1991.
The time is determined as in Fig.~\ref{fig:lc1}. The horizontal band
represents the 68\% confidence-level EGRET off-occultation flux determined
from the two-week period. Black (thin) error bars give the average flux
during the occultation seen by EGRET; red (thick) error bars are expected
for a Fermi observation, in the pointing mode, of one occultation with the
same parameters.}
\end{center}
\end{figure}

We performed the fit for the
occultation period and for the rest of the viewing period.
The best fits give the
signal of
$N_s=4.82$ photons for the occultation and $N_s=284.5$ for the rest
of the viewing period. Within the 68\% containment angle of the PSF, the
numbers of signal and background photons are roughly equal. Therefore we
found some weak evidence for a non-zero point-source flux from the
location of 3C~279 during the occultation, $(6.2^{+3.7}_{-2.7})\cdot
10^{-7}$~cm$^{-2}$~s$^{-1}$, to be compared with the value obtained from
the analysis of the rest of the same viewing period, $(8.6 \pm 0.5 )\cdot
10^{-7}$~cm$^{-2}$~s$^{-1}$ (the value quoted in the 3EG catalog for this
period is $(7.94 \pm 0.75 )\cdot 10^{-7}$~cm$^{-2}$~s$^{-1}$). The zero
flux (point-like source) is excluded at $\approx 98\%$ confidence level
(CL); no upper limit can be placed on the source extension because
unsuppressed flux is well within the 68\% CL error bars.
Assuming a Gaussian extended image, the best fit for the radial
extension is $\sigma\approx 0.283^\circ$. {The
error bars in Fig.}~\ref{fig:lc2}
give an idea of the size of statistical uncertainties both of this result
and of potential future studies.

Here we mention some possible subtleties which one should be aware of when trying to understand this result or similar future observations.

{\bf Other nearby 3EG sources.}
The numbers quoted above were obtained with the subtraction of the
expected contribution of a single nearby source detected at 3~sigma in the
v.p.~11.0 (3EG~J1235$+$0233, see the list of nearby sources in
Table~\ref{table:nearby}).
\begin{table}
\begin{center}
\begin{tabular}{ccccc}
\hline
3EG  & other & $\theta$ & $F_{-7}$             &$(TS)^{1/2}$\\
name & name  & & & \\
\hline
1219$-$1520  &          &13.2$^\circ$ & $<$0.89     & 0.0\\
1229$+$0210  &3C~273    &10.5$^\circ$ & $<$0.95     & 0.4\\
1230$-$0247  &1229$-$021& 7.0$^\circ$ &1.13$\pm$0.43& 2.9\\
1234$-$1318  &          & 9.3$^\circ$ &$<$0.89      & 0.0\\
1235$+$0233  &          & 9.9$^\circ$ &1.24$\pm$0.41& 3.5\\
1236$+$0457  &          &11.9$^\circ$ &$<$0.90      & 0.3\\
1246$-$0651  &1243$-$072& 2.5$^\circ$ &1.29$\pm$0.54& 2.7\\
1310$-$0517  &          & 3.6$^\circ$ &1.05$\pm$0.51& 2.2\\
1339$-$1419  &1334$-$127&13.6$^\circ$ &$<$1.08      & 0.0\\
\hline
1255$-$0549  &3C~279    & 0.0$^\circ$ &7.94$\pm$0.75&15.1\\
\hline
\end{tabular}
\end{center}
\caption{\label{table:nearby}Potential confusing sources: 3EG sources
within 15$^\circ$ of 3C~279. $\theta$ is the angular offset from 3C~279;
$F_{-7}$ is the flux during v.p.~11.0 in  $\rm 10^{-7}[cm^{-2}\,s^{-1}]$,
$(TS)^{1/2}$ is the significance of detection in the v.p.~11.0. Data from
\citet{3EG}.}
\end{table}
The result can in principle be confused by
contribution of other sources.
As a test, we changed the 3-sigma threshold adopted by EGRET to
both 2~sigma and to 4~sigma
without any significant change in the result.
The procedure described above assumed that the flux of the sources being
subtracted was constant during the viewing period. Clearly, an extreme
flare of 3EG~J1246$-$0651, or 3EG~J1310$-$0517, or both, exactly at the
occultation time, could explain our result without 3C~279. Note however
that on the time scale between one viewing period and another, these
sources do not demonstrate significant variability so such a flare seems
unlikely.

{\bf The Sun.}
The solar surface could be a gamma-ray source due to its
interaction with cosmic rays. Early EGRET studies put
a 95\%~C.L.\ upper limit of $2.0\cdot
10^{-7}$~cm$^{-2}$~s$^{-1}$ on the flux of the quiet
Sun~\citep{EGRET:SunMoon}. A marginal detection of solar disk flux
$\sim 4\cdot 10^{-8}$~cm$^{-2}$~s$^{-1}$ has been reported by
\citet{newOrlStr}. The theoretical expectation of the flux of the disk of
the quiet Sun is about $2\cdot
10^{-8}$~cm$^{-2}$~s$^{-1}$~\citep{Sunextended3}.

The extended emission of
the Sun
\citep{Sunextended1,Sunextended2,newOrlStr} cannot explain the observed
excess: within the 68\% containment width of the EGRET PSF the
expected flux from the extended solar emission is about $\sim
1.5\times 10^{-7}$~cm$^{-2}$~s$^{-1}$ assuming the model for the solar
extended flux and the total flux of $4.44\cdot 10^{-7}$~cm$^{-2}$~s$^{-1}$
obtained by \citet{newOrlStr} in the 10-degree circle around the Sun. We
would like to note that the recent unpublished Fermi observations have
been used to extract the solar flux from the $10^\circ$ region around the
Sun which is $(4.59\pm0.89)\cdot 10^{-7}$~cm$^{-2}$~s$^{-1}$
\citep{Brigida:2009}, in a good egreement with \citet{newOrlStr}.
This flux is low compared to that of 3C~379 and
may be neglected within our current poor precision.

{\bf Solar flares.} Solar flares are sources of gamma
rays; the BATSE records ({\tt
http://umbra.nascom.nasa.gov/batse/batse\underline{~}years.html}) list
four weak flares in the occultation time (see Table~\ref{table:flares}).
\begin{table}
\begin{center}
\begin{tabular}{ccc}
\hline
Start, UT & End, UT & EGRET photons\\
& &(separation from the Sun)\\
\hline
20$^{\rm h}$14$^{\rm m}$ & 20$^{\rm h}$15$^{\rm m}$ & 27.2$^\circ$,
32.3$^\circ$, 21.7$^\circ$, 19.4$^\circ$, 17.8$^\circ$\\
20$^{\rm h}$48$^{\rm m}$ & 20$^{\rm h}$49$^{\rm m}$ & --\\
21$^{\rm h}$59$^{\rm m}$ & 22$^{\rm h}$00$^{\rm m}$ & 10.8$^\circ$\\
22$^{\rm h}$07$^{\rm m}$ & 22$^{\rm h}$08$^{\rm m}$ & --\\
\hline
\end{tabular}
\end{center}
\caption{\label{table:flares} Solar flares during the occultation.}
\end{table}
The
photons detected by EGRET during these flares were separated from the Sun
by at least $10.8^\circ$, so they most probably do not contribute to
the point-source flux (the 68\% width of the EGRET PSF at $E>100$~MeV is
$3.3^\circ$).

{\bf The Moon.}
One more
gamma-ray source was nearby during the time of these EGRET observations:
the Moon was in $6^\circ$ to $9^\circ$ from the Sun during the period of
interest. The gamma-ray ($E>100$~MeV) flux of the Moon in 1991 was  $(3.6
\pm 0.9 )\cdot 10^{-7}$~cm$^{-2}$~s$^{-1}$ (see Fig.~2 of
\citet{EGRET:SunMoon}). Given the separation, the PSF width and the flux we
conclude that the lunar contribution cannot explain the observed excess.

{\bf Possibility of misidentification.}
Though 3C~279 is considered as one of the best EGRET identifications, one
still cannot exclude the possibility that the gamma-ray excess is due to
a source mis-identified with 3C~279. This would-be actual source, if
located just 20 arc min away, would not be screened by the Sun.
The best-fit position of the EGRET source associated with
3C~279 is indeed displaced from the position of the quasar as seen in other wavelengths, but this position is evenly more deeply screened by the Sun during an occultation.

\section{Prospects of future observations}
\label{sec:GLAST}
 Let us estimate the ability of new experiments to observe
the solar
 occultations of gamma-ray sources, notably that of 3C~279. AGILE
cannot be
 pointed to the Sun because of configuration of its solar panels
(M.~Tavani, private communication). Atmospheric Cerenkov telescopes cannot
be pointed at the Sun as they would be destroyed. The sensitivity of MILAGRO
is insufficient
 to detect the source in 8.5 hours. The Large Area Telescope
(LAT) of Fermi
 may however be used to observe the occultation succesfully.

 In the survey mode, Fermi will scan the sky rotating continuously, so the
short-period exposure to a given point in the sky is not too large. The
sensitivity may be estimated using the Fermi web service
(http://www-glast.slac.stanford.edu/software/IS/glast{\underline{~}}lat{\underline{~}}performance.htm).
For the pointing mode, the average on-axis effective area for $E> 100$~MeV, weighted with
the $\sim E^{-2}$ spectrum, 
 is $\sim 4700~{\rm cm}^2$.
Further precision may  be gained  by repeating  the
observation each year.

\begin{figure}
\begin{center}
\includegraphics[width=8.3cm]{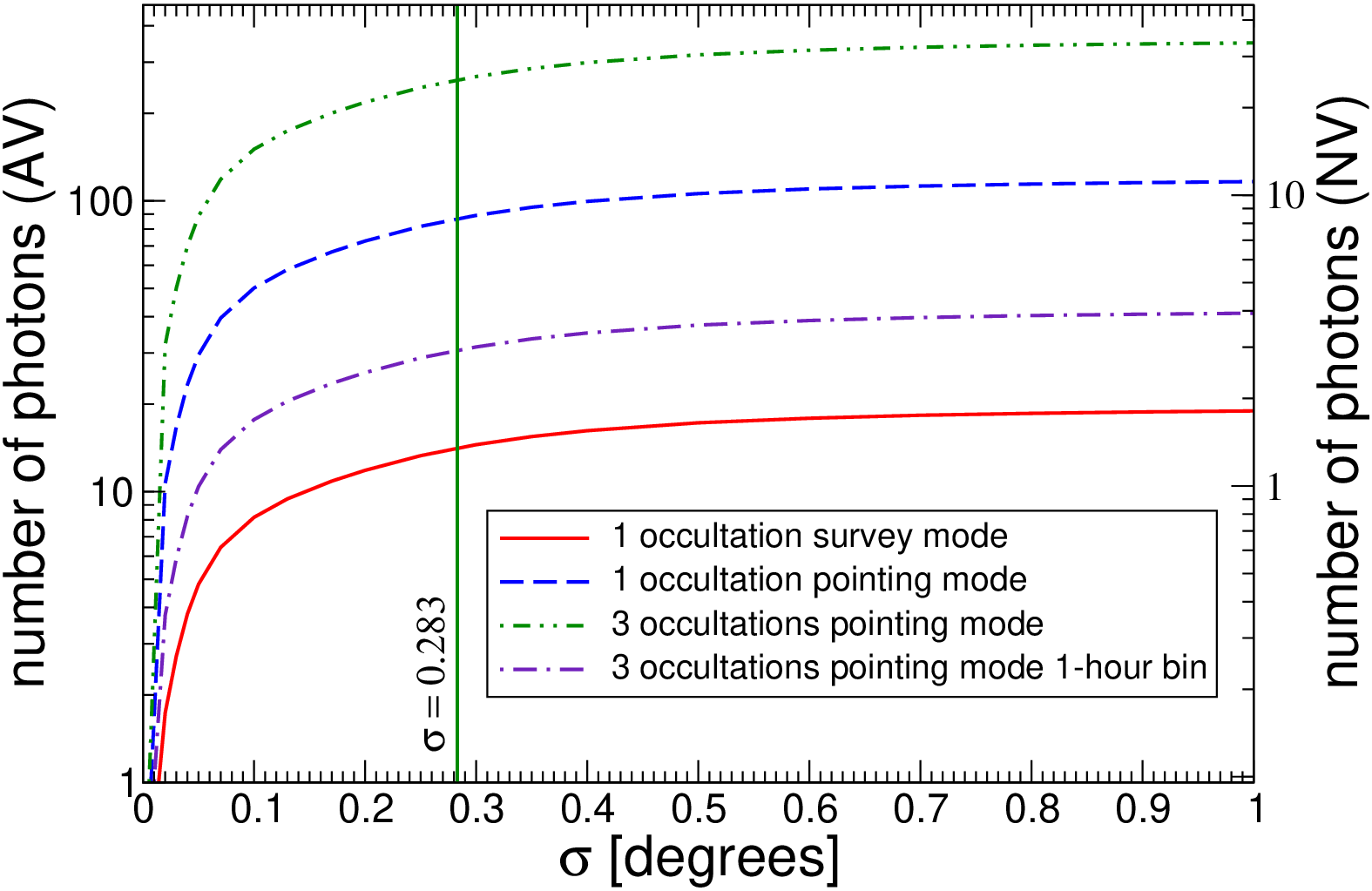}
\caption{
\label{fig:photons}
The expected number of source photons from 3C~279 during the occultation,
seen by Fermi, versus the radial extent of the Gaussian image, for
8.5--hour observations in the survey mode (full line), pointing mode
(dashed line), combination of three observations in the pointing mode
(dash--dot--dotted line) and for 1--hour bin of the latter combination
(dash-dotted line). The vertical line indicates the best-fit value of the
extension from the EGRET data.
The left and right scales are described in the text.
}
\end{center}
\end{figure}

To estimate the ability of the instrument to detect non-zero flux during
the occultation, we have to assume a particular value of the total flux of
the source. 3C~279 is strongly variable; however the extended image should
not be variable unless it is formed near the Earth and not near the
quasar. While formation of the halo near the quasar is better motivated
physically (cf.\ Sec.~\ref{sec:extended}), the opposite is in better
agreement with the EGRET result (Sec.~\ref{sec:EGRET}) because the
non-variable flux of 3C~279, roughly estimated as a minimal flux over
EGRET viewing periods, is $\sim 8 \times 10^{-8}$~cm$^{-2}$s$^{-1}$
\citep{3EG}, much lower than the best-fit flux during the occultation.
Figure~\ref{fig:photons}
shows the expected number of photons from the
source during the occultation as a function of the source extension for
various Fermi observations for both scenarios.  The left-handed scale (AV)
corresponds to the total flux of the Gaussian image of $83.7 \times
10^{-8}$~cm$^{-2}$s$^{-1}$, the average flux of 3C~279 over nine viewing
periods \citep{Grenier}; the right-handed scale (NV) corresponds to the
non-variable flux.  The number of background photons within the 68\%
containment width of PSF is approximately equal to the number of source
photons in the latter scenario.

Two more EGRET sources with ecliptic latitudes
$b_e<0.25^\circ$ (see Table~\ref{tab:other})
\begin{table}
\begin{center}
\begin{tabular}{ccccc}
\hline
name & $b_e$ &3EG & GEV & VHE\\
\hline
AX~J1809.8$-$2333 & $-$0.099 &J1809$-$2328 & yes & no\\
3C~279            & $+$0.186 &J1255$-$0549 & yes & MAGIC\\
W28$=$M20 (?)     & $-$0.020 &J1800$-$2338 & yes & HESS\\
\hline
\end{tabular}
\end{center}
\caption{\label{tab:other} (EGRET and FERMI gamma-ray sources potentially eclipsed by the Sun. $b_e$
  gives the ecliptic latitude calculated from the FERMI bright source list~\citep{Abdo:2009mg}, ``GEV'' means $E>1$~GeV detection by EGRET~\citep{GeV},
  ``VHE'' means $E>100$~GeV detection. }
\end{table}
had been
classified as
unidentified in the 3EG catalog. Further studies suggested potential
identifications; they also have been detected by FERMI~\citep{Abdo:2009mg}
The study of the solar occultation
may help to determine their coordinates with higher precision, testing
the
identification. We expect that these (and maybe other)
sources will
become
potential targets for angular-size measurements.

\section{Conclusion}
\label{sec:concl}
It will be interesting to try and measure the angular sizes of images of
energetic gamma--ray sources by means of observation of their solar
occultations. The best target is 3C~279, whose occultation happens each
year on the 8th of October.  EGRET observations made during such a period did not
exclude the unsuppressed flux of the quasar when it was screened by the
Sun. The sensitivity of the Fermi telescope is high enough that if the
flux was unsuppressed during occultation, it could be observed more
definitively than with EGRET.  Fermi can also constrain the angular size
of the image even in the survey mode, and is capable of  obtaining a light
curve by the combination of several observations in the pointing mode.
This would help to constrain models of particle acceleration and magnetic
fields in and around the quasar. If the flux during the occultation exceeds
the non-variable flux of the source (as it is slightly favoured by the
EGRET data), it would mean that either the extended image is formed
relatively nearby or the Sun is partially transparent for the point-like
gamma-ray emission (both options would mean a discovery of some
unconventional physical or astrophysical phenomenon). The same method may
be applied to refine the coordinates and/or to estimate the angular size
of images of other gamma-ray sources screened by the Sun.

We are grateful for discussions with J.~Conrad,
V.~Rubakov and M.~Tavani.  This work was supported in part by DFG (Germany) and
CONACYT (Mexico) (TR), by the grants RFBR 07-02-00820, RFBR 09-07-00388,
NS-1616.2008.2 and by FASI under state contracts 02.740.11.0244 and
02.740.11.5092 (ST). We made use of NED which is operated by the Jet
Propulsion Laboratory, CalTech, under contract with NASA.


\bsp

\label{lastpage}

\end{document}